\documentclass{article}
\usepackage{spconf,amsmath,graphicx,hyperref}
\usepackage{tabularx}
\usepackage{booktabs}
\usepackage{pgfplots}
\pgfplotsset{compat=1.18} 
\usepackage{array} 
\usepackage{graphicx}
\usepackage{booktabs}
\usepackage{amsfonts}
\usepackage{makecell} 
\usepackage{multirow} 
\usepackage{arydshln}
\usepackage{cite}  
\usepackage{tikz}
\usepackage{xcolor}
\definecolor{deepPurple}{rgb}{0.267,0.004,0.329}  
\definecolor{myBlue}{rgb}{0,0,1}                  
\definecolor{myGreen}{rgb}{0,0.502,0}             
\definecolor{myOrange}{rgb}{1,0.647,0}            
\definecolor{myYellow}{rgb}{1,1,0}                

\title{From Human Speech to Ocean Signals: Transferring Speech Large Models for Underwater Acoustic Target Recognition}
\name{
Mengcheng Huang,
Xue Zhou,
Chen Xu\textsuperscript{*}\thanks{Corresponding author: chen.xu@hrbeu.edu.cn.},
Dapeng Man
}
\address{College of Computer Science and Technology, Harbin Engineering University, Harbin, China}
\begin{document}
\ninept
\maketitle
\begin{abstract}
Underwater acoustic target recognition (UATR) plays a vital role in marine applications but remains challenging due to limited labeled data and the complexity of ocean environments. This paper explores a central question: can speech large models (SLMs), trained on massive human speech corpora, be effectively transferred to underwater acoustics? 
To investigate this, we propose UATR-SLM, a simple framework that reuses the speech feature pipeline, adapts the SLM as an acoustic encoder, and adds a lightweight classifier.
Experiments on the DeepShip and ShipsEar benchmarks show that UATR-SLM achieves over 99\% in-domain accuracy, maintains strong robustness across variable signal lengths, and reaches up to 96.67\% accuracy in cross-domain evaluation. 
These results highlight the strong transferability of SLMs to UATR, establishing a promising paradigm for leveraging speech foundation models in underwater acoustics.
\end{abstract}
\begin{keywords}
Underwater Acoustic Target Recognition, Speech Large Models, Transfer Learning, Cross-Domain Generalization
\end{keywords}
\section{Introduction}
\label{sec:intro}

Underwater acoustic target recognition (UATR) is fundamental to marine research and operations, supporting applications such as biodiversity monitoring, search and rescue, seabed mapping, and maritime security \cite{xie2024unraveling,zhao2025enhancing}. 
Despite its importance, developing reliable UATR systems remains challenging due to two persistent issues: 
(1) \textbf{data scarcity}, as collecting and annotating large-scale underwater acoustic datasets is expensive and labor-intensive \cite{tian2023few}, and (2) \textbf{domain complexity}, since underwater environments exhibit nonstationary noise, multipath propagation, and substantial spatiotemporal variability \cite{feng2024artificial}.
These factors limit the performance and robustness of conventional deep learning methods.

\begin{figure}[htbp]
    \centering
    \includegraphics[width=0.55\linewidth]{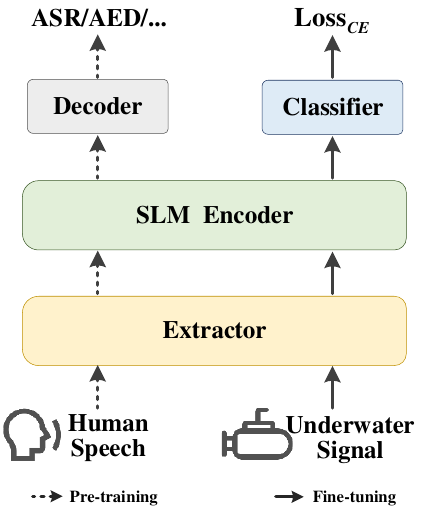}
    \caption{Overview of UATR-SLM. Underwater signals follow the standard speech feature pipeline, are encoded by a speech large model pretrained on massive corpora, and are classified by a lightweight head.}
    \label{fig:model}
\end{figure}

Existing research tackles these issues through data-centric and model-centric strategies \cite{domingos2022survey}. To alleviate data scarcity, researchers have explored data augmentation (e.g., spectrogram masking \cite{park2019specaugment}, pitch shifting \cite{zhang2021pdaugment}, and generative synthesis \cite{qian2019data}) and transfer learning from related domains such as environmental sound recognition. To handle domain complexity, approaches such as noise-robust feature extraction, domain adaptation, and adversarial training have been introduced to improve generalization across environments \cite{wang2018supervised}. 
While these methods provide incremental improvements, they remain fundamentally constrained by the limited scale and diversity of existing pretraining data, and they often fail to capture universal acoustic patterns that generalize well across environments.

Recent advances in foundation models suggest a potential solution. \emph{Speech large models} (SLMs), trained on massive human speech corpora, have demonstrated remarkable generalization ability in diverse speech and audio tasks \cite{fang2024llama}. 
While human speech and underwater acoustics are physically distinct domains, they share low-level structures such as spectral patterns and temporal dynamics. 
This motivates a central question: 
\emph{to what extent can acoustic knowledge captured by SLMs transfer to underwater acoustics, and how effectively can it be adapted for UATR?}
Addressing this question could pave the way toward a new paradigm in which general-purpose audio modeling capabilities alleviate both data scarcity and domain variability in underwater tasks.

In this work, we introduce \textbf{UATR-SLM}, a simple yet effective framework that adapts SLMs for underwater acoustic target recognition, as shown in Figure \ref{fig:model}. 
The core idea is to leverage the SLM as an acoustic encoder, process underwater signals through the same feature extraction pipeline as speech, and then apply a lightweight classifier for target recognition.
Importantly, this design requires only a small amount of task-specific fine-tuning, making it suitable for data-scarce underwater settings.

We validate UATR-SLM through extensive experiments on two widely used benchmarks: DeepShip and ShipsEar. Results show that our model consistently achieves over 99\% accuracy on both datasets, demonstrates strong robustness across variable-length signals, and exhibits remarkable cross-domain generalization, achieving 96.67\% accuracy when transferring to unseen environments. 
These findings establish that large speech models, despite being trained exclusively on human speech, encode rich acoustic priors that can be repurposed for non-speech domains such as underwater acoustics.

Our main contributions are summarized as follows:
\begin{itemize}
    \item 
    To the best of our knowledge, we present the first systematic study of transferring  speech large models to underwater acoustic target recognition.
    \item 
    We propose UATR-SLM, a simple and effective framework that leverages SLMs as universal acoustic encoders and enables efficient fine-tuning, thereby addressing the challenge of data scarcity.
    \item 
    Through extensive experiments, we demonstrate that UATR-SLM achieves state-of-the-art performance, robustness across signal lengths, and promising cross-domain generalization.
\end{itemize}

These findings highlight the potential of speech large models as highly transferable acoustic encoders, enabling efficient adaptation from human speech to underwater acoustics. 
As future work, we aim to further explore lightweight and deployable variants of UATR-SLM for enabling practical applications in real-world marine environments.

\section{Method}
\label{sec:format}

\subsection{Problem Formulation}

Underwater acoustic target recognition can be naturally formulated as a supervised classification problem \cite{luo2023survey}.
Given a raw underwater acoustic signal $x \in \mathbb{R}^T$ of length $T$, the goal is to predict its class label $y \in \{1, \dots, C\}$, where $C$ denotes the number of target categories (e.g., different ship types). 
The difficulty of this task stems from two major factors: (1) the scarcity of labeled underwater acoustic data, and (2) the substantial domain variability introduced by diverse environmental conditions. 
These issues motivate transferring knowledge from speech large models, which have already learned general acoustic structures from massive corpora.

\subsection{Progress of Speech Large Models}

Self-supervised learning has driven remarkable progress in speech representation learning. 
Early models such as wav2vec 2.0 \cite{baevski2020wav2vec}, HuBERT \cite{hsu2021hubert}, and WavLM \cite{chen2022wavlm} showed that large-scale training on raw waveforms or spectrograms can capture phonetic, prosodic, and temporal cues without requiring labeled data. 
Building on this foundation, large-scale models such as Whisper \cite{radford2023robust}, SeamlessM4T \cite{barrault2023seamlessm4t}, and SenseVoice \cite{an2024funaudiollm} have expanded into the era of speech foundation models, delivering strong performance across a wide range of tasks, including automatic speech recognition, spoken question answering, and speaker verification.

A key characteristic of SLMs is their universality. Although trained primarily on human speech, these models learn acoustic representations that are not limited to linguistic content. 
Prior studies have shown successful transfer of SLMs to paralinguistic analysis, emotion recognition, and even environmental sound classification \cite{baevski2020wav2vec,hsu2021hubert,chen2022wavlm,radford2023robust,barrault2023seamlessm4t,an2024funaudiollm}, suggesting that they capture generalizable spectral and temporal regularities.

Despite this progress, applying SLMs to non-speech acoustic domains remains underexplored. 
Underwater acoustics provides a particularly challenging case: it differs fundamentally from human speech in its physical properties, yet still shares low-level spectral and temporal structures. 
This gap motivates our central question: to what extent can SLMs transfer to underwater acoustics, and how effectively can they be adapted for UATR?

\subsection{UATR-SLM Framework}

To answer this question, we propose UATR-SLM, a simple yet effective framework for adapting SLMs to underwater acoustic target recognition. 
As illustrated in Figure~\ref{fig:model}, the framework consists of three main components:

\begin{itemize}
    \item \textbf{Feature Extractor.} Raw underwater acoustic signals are processed using the same preprocessing pipeline employed for speech in SLMs. For example, log-Mel filterbank features are extracted, consecutive frames are stacked, and the sequence is downsampled to match the input format of the SLM. This alignment ensures compatibility and allows direct reuse of pretrained models.
    
    \item \textbf{Encoder.} The processed spectrograms are then passed into the SLM encoder, which acts as a universal acoustic model. Unlike conventional transfer learning strategies that freeze most encoder layers, we allow full fine-tuning of the encoder to account for the substantial distributional differences between human speech and underwater acoustics. This strategy enables the model to adapt more comprehensively to the unique characteristics of underwater signals.


    \item \textbf{Lightweight Classifier.} Most SLMs are designed for generative tasks and include a decoder with large vocabularies, which is not directly applicable to classification tasks. To address this, we replace the decoder with a lightweight classifier head: the encoder outputs are aggregated using mean pooling, followed by a single linear layer and a Softmax function that maps features to probabilities over $C$ target categories. 

\end{itemize}

Together, these three components enable the effective transfer of SLMs to UATR.
The framework leverages the powerful representation learning capacity of speech models while introducing only minimal task-specific modifications, making it both simple and efficient for deployment in data-scarce underwater environments.
Although we have conducted preliminary ablation studies to validate these design choices, due to space constraints we do not report them here, and instead focus on presenting the overall superior performance of UATR-SLM.

\section{Experiment}
\label{sec:Experiment}

\subsection{Experimental Settings}

\subsubsection{Datasets and Pre-processing}
\label{subsubsec:Datasets and Pre-processing}

To evaluate the UATR-SLM framework, we conduct experiments on two widely used underwater acoustic benchmarks benchmarks: DeepShip \cite{irfan2021deepship} and ShipsEar \cite{santos2016shipsear}. 
DeepShip is a large-scale dataset which collected in the Georgia Strait, and comprises recordings from 265 vessels across four categories: cargo ships, passenger ships, oil tankers, and tugs. 
In contrast, ShipsEar is a smaller and more challenging dataset containing radiated noise from 11 vessel types, along with ambient background noise. 
Following prior work, we consolidate the 11 vessel types into four broader categories and retain the background class, resulting in a five-class classification task.

A unified preprocessing pipeline was applied to both datasets. All recordings were resampled to 16 kHz and segmented into non-overlapping 5-second clips. 
This produced over 30,000 samples for DeepShip and 2,223 samples for ShipsEar. 
For all experiments, we adopt a standard 8:1:1 split for train/validation/test sets. 
The detailed dataset statistics are reported in Table~\ref{tab:deepship} and Table~\ref{tab:shipsear}.

\begin{table}[htbp]
\centering
\caption{Statistics of the DeepShip dataset.}
\label{tab:deepship}
\begin{tabular}{lrrrr}
\toprule
Class & Train & Validation & Test & Total \\ 
\midrule
Cargo       & 6,097    & 762        & 762     & 7,621 \\
Passenger   & 7,369    & 921        & 921     & 9,211 \\
Tanker      & 7,022    & 877        & 877     & 8,776 \\
Tug         & 6,467    & 809        & 809     & 8,085 \\
\midrule
\textbf{Total} & 26,955   & 3,369      & 3,369   & 33,693 \\
\bottomrule
\end{tabular}
\end{table}

\begin{table}[htbp]
\centering
\caption{Statistics of the ShipsEar dataset.}
\label{tab:shipsear}
\begin{tabular}{lrrrr}
\toprule
Class & Train & Validation & Test & Total \\ 
\midrule
Passenger & 675 & 84 & 84 & 843 \\ 
\specialrule{0em}{0.4pt}{0.4pt}
\cdashline{1-5}[2pt/3pt]
\specialrule{0em}{0.5pt}{1pt}
\makecell[l]{Ocean liner, RoRo} & 389 & 49 & 48 & 486 \\
\specialrule{0em}{0.4pt}{0.4pt}
\cdashline{1-5}[2pt/3pt]
\specialrule{0em}{0.5pt}{1pt}
\makecell[l]{Fish boat, Trawler, \\ Tugboat, Dredger, \\Mussel boat} & 296 & 37 & 36 & 369 \\
\specialrule{0em}{0.4pt}{0.4pt}
\cdashline{1-5}[2pt/3pt]
\specialrule{0em}{0.5pt}{1pt}
\makecell[l]{Pilot boat, \\Sailboat, Motorboat} & 241 & 30 & 30 & 301 \\ 
\specialrule{0em}{0.4pt}{0.4pt}
\cdashline{1-5}[2pt/3pt]
\specialrule{0em}{0.5pt}{1pt}
Background noise & 180 & 22 & 22 & 224 \\ 
\midrule
\textbf{Total} & 1,781 & 222 & 220 & 2,223 \\
\bottomrule
\end{tabular}
\end{table}

\subsubsection{Model Settings}
\label{subsubsec:Model Settings}


In this work, we adopt SenseVoiceSmall, a versatile speech foundation model designed for low-latency multilingual processing, as the encoder in our framework for three main reasons:
(1) Universal modeling capability: it supports a wide range of speech understanding tasks, including automatic speech recognition (ASR), spoken language identification (LID), speech emotion recognition (SER), and audio event detection (AED), indicating strong potential to capture the complex spectral and temporal patterns of underwater acoustics;
(2) Parameter efficiency: trained on 300,000 hours of multilingual and multi-task speech data yet containing only 234M parameters, it is lightweight and well-suited for UATR under low-latency, resource-constrained settings;
(3) Architecture adaptability: as an encoder-only model with a CTC-based \cite{graves2006connectionist} ASR module, its modeling capacity is fully concentrated in the encoder, enabling direct and comprehensive transfer of acoustic knowledge to the UATR task.

For fine-tuning, we adjust hyperparameters according to dataset scale: a batch size of 60 with a learning rate of 2e-4 on DeepShip, and a batch size of 10 with a learning rate of 4e-5 on ShipsEar. All experiments use the AdamW optimizer \cite{loshchilov2017decoupled}, a WarmupLR scheduler, and cross-entropy loss.

\subsubsection{Evaluation Settings}

To comprehensively assess the effectiveness of UATR-SLM on underwater acoustics, we design three complementary evaluations:
\begin{itemize}

    \item \textbf{Basic performance}: We first evaluate the in-domain classification accuracy of UATR-SLM on the DeepShip and ShipsEar datasets. This directly measures how well the model performs when trained and tested within the same distribution, serving as the foundation for subsequent analyses.

    \item \textbf{Robustness}: We then evaluate the model’s stability under challenging conditions by testing recognition performance across variable signal lengths. This setting examines temporal robustness, a key requirement for real-world deployments where recordings may be short and noisy.
    \item \textbf{Cross-domain generalization}: Finally, we investigate the model’s ability to generalize across datasets under zero-shot conditions. Specifically, we train UATR-SLM on DeepShip and evaluate on ShipsEar without fine-tuning. This reflects practical scenarios in which models must operate reliably in unseen environments without access to labeled data.
\end{itemize}

Across all evaluations, we compare UATR-SLM with classical ResNet \cite{he2016deep} baselines and report results using multiple metrics, including accuracy, precision, recall, and F1-score.

\subsection{Experimental Results}
\label{subsec:Experimental Results}

\subsubsection{Basic Performance}
\label{subsubsec:Basic Performance}

We evaluate the basic performance of UATR-SLM on the DeepShip and ShipsEar datasets, with results summarized in Tables \ref{tab:deepship_result} and \ref{tab:shipsear_result}. Across both benchmarks, UATR-SLM consistently achieves the best results on all metrics, establishing new state-of-the-art performance without requiring complex architectural designs. 


On DeepShip, our model attains a 99.32\% F1-score, surpassing all baselines and showing that adapting pretrained speech models yields more robust gains than domain-specific innovations. 
On the smaller, more challenging ShipsEar, UATR-SLM achieves a 99.00\% F1-score, outperforming ResNet baselines and recent advanced models such as SSA-CACNN, HUAT, and MobileViT.


These results demonstrate that speech large models provide highly expressive acoustic representations that can be effectively adapted to underwater target recognition. 
By fine-tuning on the exact same underwater training data as the baselines, UATR-SLM achieves outstanding in-domain performance, demonstrating the potential of SLMs as universal encoders for non-speech acoustic recognition tasks.

\begin{table}[t!]
\centering
\caption{Results of different models on the DeepShip dataset.}
\label{tab:deepship_result}
\begin{tabular}{lcccc}
\toprule
Model & Accuracy  & Precision & Recall & F1 \\
\midrule
ResNet18 & 95.90 & 95.96 & 95.81 & 95.87 \\
ResNet34 & 95.67 & 95.63 & 95.67 & 95.65 \\
ResNet50 & 92.47 & 92.37 & 92.43 & 92.39 \\
\specialrule{0em}{0.4pt}{0.4pt}
\cdashline{1-5}[2pt/3pt]
\specialrule{0em}{0.5pt}{1pt}
HUAT \cite{chen2024ship} & 99.01 & 99.01 & 99.01 & 99.01 \\
BAHTNet \cite{zhao2025enhancing} & 94.57 & 94.54 & 94.58 & 94.56 \\
SSA-CACNN \cite{ji2025underwater} & 94.76 & 95.17 & 94.76 & 94.89 \\
\midrule
UATR-SLM (Ours) & \textbf{99.32} & \textbf{99.31} & \textbf{99.32} & \textbf{99.32} \\
\bottomrule
\end{tabular}
\end{table}

\begin{table}[t!]
\centering
\caption{Results of different models on the ShipsEar dataset.}
\label{tab:shipsear_result}
\begin{tabular}{lcccc}
\toprule
Model & Accuracy & Precision & Recall & F1 \\
\midrule
ResNet18 & 96.82 & 97.06 & 96.62 & 96.77 \\
ResNet34 & 96.36 & 96.83 & 96.06 & 96.32 \\
ResNet50 & 94.55 & 95.00 & 95.47 & 95.21 \\
\specialrule{0em}{0.4pt}{0.4pt}
\cdashline{1-5}[2pt/3pt]
\specialrule{0em}{0.5pt}{1pt}
HUAT \cite{chen2024ship} & 98.62 & 98.18 & 98.82 & 98.50 \\
Mobile\_ViT \cite{yao2024mobile_vit} & 98.50 & 98.37 & 98.40 & 98.38 \\
SSA-CACNN \cite{ji2025underwater} & 98.68 & 98.52 & 98.68 & 98.59 \\
\midrule
UATR-SLM (Ours) & \textbf{99.09} & \textbf{98.80} & \textbf{99.21} & \textbf{99.00} \\
\bottomrule
\end{tabular}
\end{table}

\subsubsection{Robustness}

To further evaluate robustness, we test the models on variable-length signals, a common scenario in real-world deployments. A truly generalizable model should not rely on fixed-length inputs but instead extract discriminative features even from brief acoustic events. To simulate this, all models are fine-tuned only on 5-second clips and then evaluated in a zero-shot manner on clips of varying durations ranging from 1 to 20 seconds, without additional training.

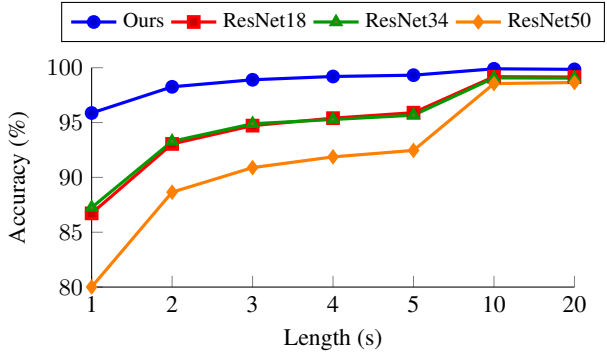
\begin{figure}[t]
    \centering
    \begin{tikzpicture}
        \begin{axis}[
            width=8cm,      
            height=4.5cm,     
            xlabel={Length (s)},
            ylabel={Accuracy (\%)},
            ymin=80, ymax=100,
            xtick=data,
            enlarge x limits=false, 
            symbolic x coords={1,2,3,4,5,10,20}, 
            ytick={80,85,90,95,100},
            legend style={at={(0.5,1.3)}, anchor=north, legend columns=-1, 
    font=\footnotesize},
            axis x line*=bottom,
            axis y line*=left,
        ]
        
        \addplot+[mark=*, color=blue, line width=1.2pt]
        coordinates {(1,95.87) (2,98.26) (3,98.90) (4,99.20) (5,99.32) (10,99.90) (20,99.85)};
        \addlegendentry{Ours}
        
        \addplot+[mark=square*, color=red, line width=1.2pt]
        coordinates {(1,86.72) (2,93.05) (3,94.72) (4,95.40) (5,95.90) (10,99.18) (20,99.15)};
        \addlegendentry{ResNet18}
        
        \addplot+[mark=triangle*, color=green!60!black, line width=1.2pt]
        coordinates {(1,87.25) (2,93.29) (3,94.89) (4,95.28) (5,95.67) (10,99.07) (20,99.05)};
        \addlegendentry{ResNet34}
        
        \addplot+[mark=diamond*, color=orange, line width=1.2pt]
        coordinates {(1,80.01) (2,88.65) (3,90.89) (4,91.87) (5,92.46) (10,98.55) (20,98.65)};
        \addlegendentry{ResNet50}
        
        \end{axis}
    \end{tikzpicture}
    \caption{Accuracy of different models on variable-length audio clips.}
    \label{fig:line_acc_updated}
\end{figure}

As shown in Figure \ref{fig:line_acc_updated}, UATR-SLM maintains strong performance across different signal lengths. Remarkably, it achieves 95.87\% accuracy from just one second of audio and quickly exceeds 99\% as the clip length increases. This immediate high accuracy demonstrates the model’s ability to leverage rich pretrained acoustic representations, enabling reliable target recognition without requiring extended temporal patterns.

In contrast, ResNet models trained from scratch exhibit higher sensitivity to signal duration. Their accuracy drops to 80.0 $\sim$ 87.3\% on one second clips, lagging behind our method by approximately 8\% to 15\%.
This difference highlights the advantage of transferring knowledge from SLMs: pretrained representations provide a powerful acoustic prior that enables effective adaptation, a capability that conventional architectures trained on limited in-domain data struggle to achieve.

\subsubsection{Cross-domain generalization}
\label{subsubsec:Cross-domain generalization}

To evaluate whether UATR-SLM learns universal acoustic representations, we designed a challenging zero-shot cross-domain experiment. Specifically, models were fine-tuned on the DeepShip dataset and directly tested on the Passenger class from ShipsEar (denoted as ShipsEar-Passenger). The ShipsEar-Passenger class contains 30 audio files of varying lengths, which were segmented into 843 clips of 5 seconds each, providing a consistent evaluation protocol. This setup measures the model’s ability to generalize knowledge to a completely unseen underwater acoustic environment.

As shown in Table~\ref{tab:cross_domain}, ResNet baselines fail to generalize, with accuracies dropping to 53.33 $\sim$ 70.00\%. 
This performance collapse reflects severe overfitting to the acoustic characteristics of DeepShip and an inability to recognize ships under domain shifts. In contrast, UATR-SLM exhibits remarkable resilience, achieving 80.31\% accuracy on 5-second clips and 96.67\% on full-length audio. These results provide direct evidence that UATR-SLM captures fundamental and invariant acoustic features, while effectively disregarding domain-specific nuisance factors that hinder conventional models.

\begin{table}[thbp!]
\centering
\caption{Zero-shot classification accuracy of different models.}
\label{tab:cross_domain}
\begin{tabular}{lcccc}
\toprule
Length & ResNet18 & ResNet34 & ResNet50 & UATR-SLM \\
\midrule
5s   & 62.28 & 59.43 & 68.20 & \textbf{80.31} \\
Full  & 60.00 & 53.33 & 70.00 & \textbf{96.67} \\
\bottomrule
\end{tabular}
\end{table}

\begin{figure}[htbp]
    \centering
    \begin{minipage}{\linewidth}
        \centering
\footnotesize
\tikz\draw[deepPurple,fill=deepPurple] (0,0) circle (2pt);\;Cargo \quad
\tikz\draw[myOrange,fill=myOrange] (0,0) circle (2pt);\;Tug \quad
\tikz\draw[myGreen,fill=myGreen] (0,0) circle (2pt);\;Tanker \quad
\tikz\draw[myBlue,fill=myBlue] (0,0) circle (2pt);\;Passenger \quad
\tikz\draw[myYellow,fill=myYellow] (0,0) circle (2pt);\;ShipsEar-Passenger
    \end{minipage}
    \begin{minipage}{0.493\linewidth}
        \centering
        \includegraphics[width=\linewidth]{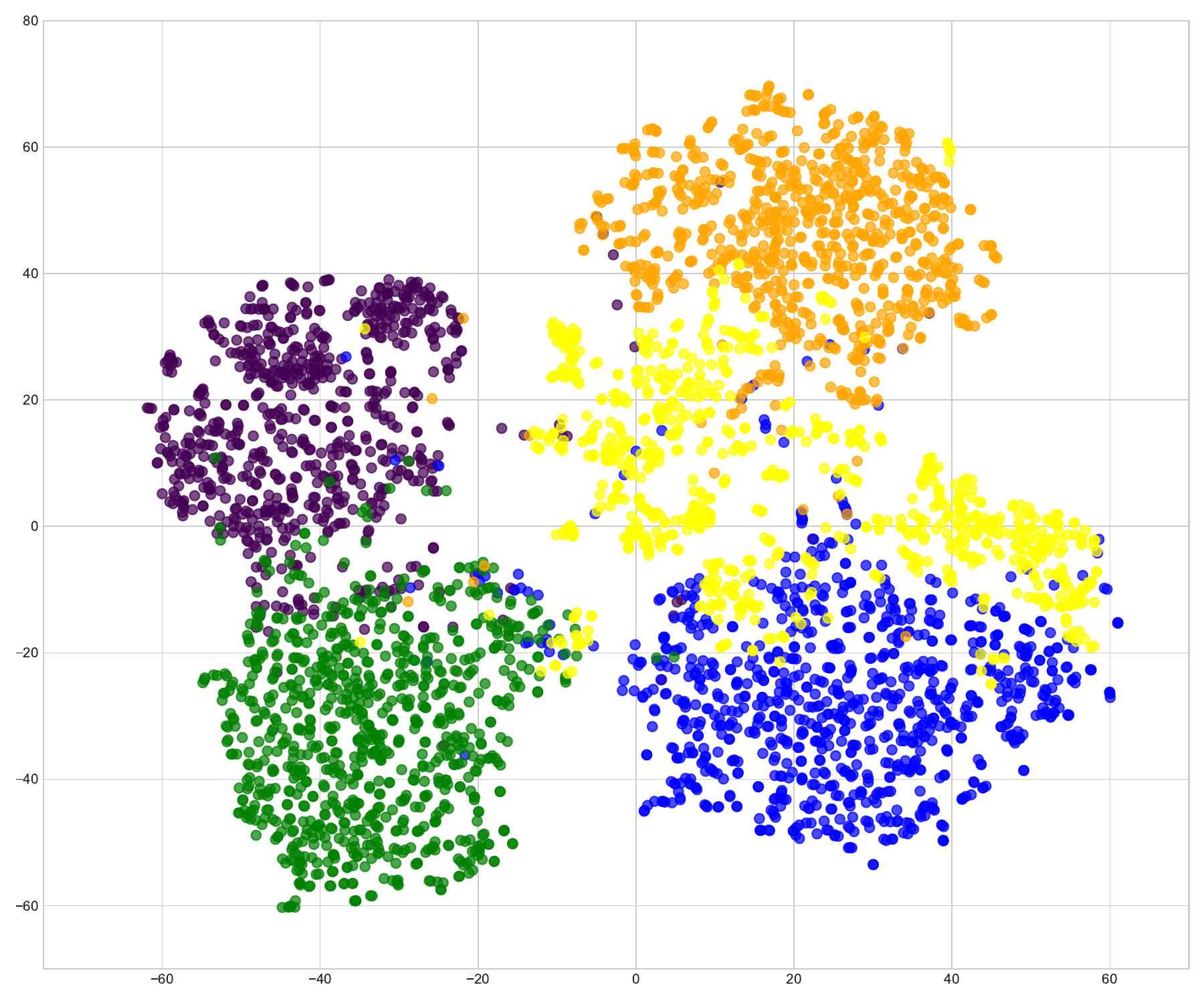}
        \caption{T-SNE of ResNet18.}
        \label{fig:T-SNE on ResNet18}
    \end{minipage}
    \begin{minipage}{0.493\linewidth}
        \centering
        \includegraphics[width=\linewidth]{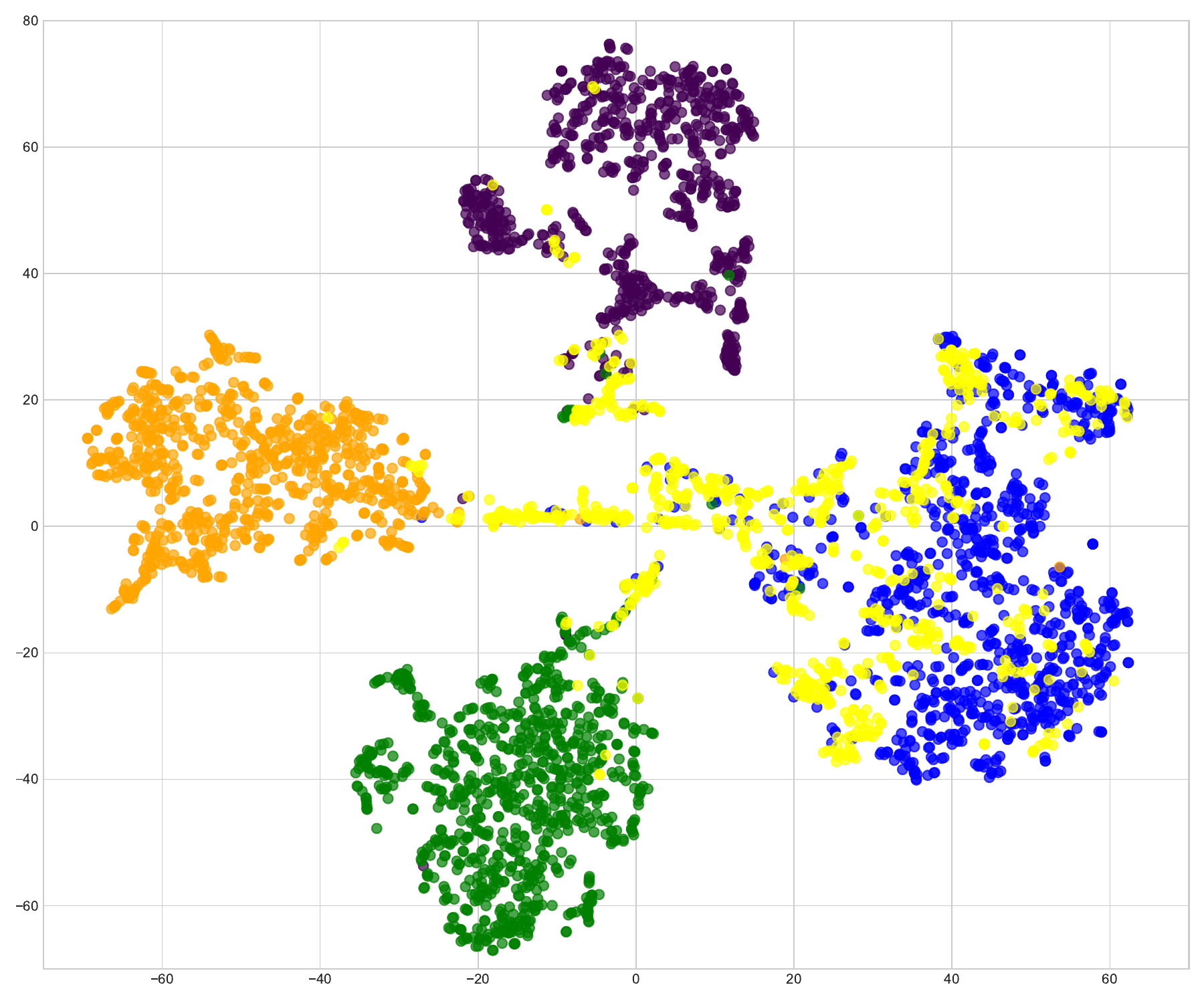}
        \caption{T-SNE of UATR-SLM.}
        \label{fig:T-SNE on UATR-SLM}
    \end{minipage}
\end{figure}


To provide an intuitive qualitative validation of the quantitative results in Table~\ref{tab:cross_domain}, we visualize the high-dimensional feature embeddings produced by ResNet18 and UATR-SLM after fine-tuning on the DeepShip dataset using T-SNE. 
The results are shown in Figures~\ref{fig:T-SNE on ResNet18} and \ref{fig:T-SNE on UATR-SLM}.

These visualizations reveal a clear difference in the quality of learned representations. In both models, the four vessel classes from the source DeepShip dataset form distinct clusters, indicating strong discriminative ability within the training domain. 
However, the critical insight comes from the mapping of the target ShipsEar-Passenger class (yellow). For UATR-SLM, these samples are positioned remarkably close to the Passenger class from DeepShip (dark blue), despite originating from an entirely unseen acoustic environment. This spatial alignment provides compelling evidence that UATR-SLM learns domain-invariant representations, capturing the intrinsic acoustic signatures of passenger vessels while suppressing domain-specific variations. In contrast, ResNet18 fails to align the unseen class effectively, highlighting its limited generalization capability.

\section{Conclusion and Future Work}
\label{sec:conclusion}

This work establishes a new paradigm for underwater acoustic target recognition by showing that speech large models can be effectively adapted to a physically distinct domain. Through UATR-SLM, we demonstrate that knowledge learned from massive speech corpora transfers remarkably well to underwater acoustics, enabling state-of-the-art performance, strong robustness to variable signal lengths, and promising cross-domain generalization. These findings highlight the potential of SLMs as universal acoustic encoders, opening a path toward more generalizable and data-efficient solutions in ocean sensing.

In future work, we will explore lightweight variants for deployment on resource-limited marine platforms, investigate self-supervised adaptation to further reduce reliance on labeled data, and extend the framework to multi-modal fusion for enhanced robustness in complex ocean conditions.

\section{Acknowledgments}
This work was supported by the National Natural Science Foundation of China (No.62406086, No.62272127) and the Joint Funds of the National Natural Science Foundation of China (No.U22A2036).

\bibliographystyle{IEEEbib}
\bibliography{strings}

@article{xie2024unraveling,
  title={Unraveling complex data diversity in underwater acoustic target recognition through convolution-based mixture of experts},
  author={Xie, Yuan and Ren, Jiawei and Xu, Ji},
  journal={Expert Systems with Applications},
  volume={249},
  pages={123431},
  year={2024},
  publisher={Elsevier}
}

@article{feng2024artificial,
  title={Artificial intelligence-based underwater acoustic target recognition: A survey},
  author={Feng, Sheng and Ma, Shuqing and Zhu, Xiaoqian and Yan, Ming},
  journal={Remote Sensing},
  volume={16},
  number={17},
  pages={3333},
  year={2024},
  publisher={MDPI}
}

@article{tian2023few,
  title={Few-shot learning for joint model in underwater acoustic target recognition},
  author={Tian, Shengzhao and Bai, Di and Zhou, Junlin and Fu, Yan and Chen, Duanbing},
  journal={Scientific Reports},
  volume={13},
  number={1},
  pages={17502},
  year={2023},
  publisher={Nature Publishing Group UK London}
}

@article{domingos2022survey,
  title={A survey of underwater acoustic data classification methods using deep learning for shoreline surveillance},
  author={Domingos, Lucas CF and Santos, Paulo E and Skelton, Phillip SM and Brinkworth, Russell SA and Sammut, Karl},
  journal={Sensors},
  volume={22},
  number={6},
  pages={2181},
  year={2022},
  publisher={MDPI}
}

@inproceedings{radford2023robust,
  title={Robust speech recognition via large-scale weak supervision},
  author={Radford, Alec and Kim, Jong Wook and Xu, Tao and Brockman, Greg and McLeavey, Christine and Sutskever, Ilya},
  booktitle={International conference on machine learning},
  pages={28492--28518},
  year={2023},
  organization={PMLR}
}

@article{fang2024llama,
  title={Llama-omni: Seamless speech interaction with large language models},
  author={Fang, Qingkai and Guo, Shoutao and Zhou, Yan and Ma, Zhengrui and Zhang, Shaolei and Feng, Yang},
  journal={arXiv preprint arXiv:2409.06666},
  year={2024}
}

@article{park2019specaugment,
  title={Specaugment: A simple data augmentation method for automatic speech recognition},
  author={Park, Daniel S and Chan, William and Zhang, Yu and Chiu, Chung-Cheng and Zoph, Barret and Cubuk, Ekin D and Le, Quoc V},
  journal={arXiv preprint arXiv:1904.08779},
  year={2019}
}

@article{barrault2023seamlessm4t,
  title={SeamlessM4T: massively multilingual \& multimodal machine translation},
  author={Barrault, Lo{\"\i}c and Chung, Yu-An and Meglioli, Mariano Cora and Dale, David and Dong, Ning and Duquenne, Paul-Ambroise and Elsahar, Hady and Gong, Hongyu and Heffernan, Kevin and Hoffman, John and others},
  journal={arXiv preprint arXiv:2308.11596},
  year={2023}
}

@article{qian2019data,
  title={Data augmentation using generative adversarial networks for robust speech recognition},
  author={Qian, Yanmin and Hu, Hu and Tan, Tian},
  journal={Speech Communication},
  volume={114},
  pages={1--9},
  year={2019},
  publisher={Elsevier}
}

@article{wang2018supervised,
  title={Supervised speech separation based on deep learning: An overview},
  author={Wang, DeLiang and Chen, Jitong},
  journal={IEEE/ACM transactions on audio, speech, and language processing},
  volume={26},
  number={10},
  pages={1702--1726},
  year={2018},
  publisher={IEEE}
}

@article{zhang2021pdaugment,
  title={Pdaugment: Data augmentation by pitch and duration adjustments for automatic lyrics transcription},
  author={Zhang, Chen and Yu, Jiaxing and Chang, LuChin and Tan, Xu and Chen, Jiawei and Qin, Tao and Zhang, Kejun},
  journal={arXiv preprint arXiv:2109.07940},
  year={2021}
}

@article{baevski2020wav2vec,
  title={wav2vec 2.0: A framework for self-supervised learning of speech representations},
  author={Baevski, Alexei and Zhou, Yuhao and Mohamed, Abdelrahman and Auli, Michael},
  journal={Advances in neural information processing systems},
  volume={33},
  pages={12449--12460},
  year={2020}
}

@article{hsu2021hubert,
  title={Hubert: Self-supervised speech representation learning by masked prediction of hidden units},
  author={Hsu, Wei-Ning and Bolte, Benjamin and Tsai, Yao-Hung Hubert and Lakhotia, Kushal and Salakhutdinov, Ruslan and Mohamed, Abdelrahman},
  journal={IEEE/ACM transactions on audio, speech, and language processing},
  volume={29},
  pages={3451--3460},
  year={2021},
  publisher={IEEE}
}

@article{chen2022wavlm,
  title={Wavlm: Large-scale self-supervised pre-training for full stack speech processing},
  author={Chen, Sanyuan and Wang, Chengyi and Chen, Zhengyang and Wu, Yu and Liu, Shujie and Chen, Zhuo and Li, Jinyu and Kanda, Naoyuki and Yoshioka, Takuya and Xiao, Xiong and others},
  journal={IEEE Journal of Selected Topics in Signal Processing},
  volume={16},
  number={6},
  pages={1505--1518},
  year={2022},
  publisher={IEEE}
}

@article{luo2023survey,
  title={A survey of underwater acoustic target recognition methods based on machine learning},
  author={Luo, Xinwei and Chen, Lu and Zhou, Hanlu and Cao, Hongli},
  journal={Journal of Marine Science and Engineering},
  volume={11},
  number={2},
  pages={384},
  year={2023},
  publisher={MDPI}
}

@article{an2024funaudiollm,
  title={Funaudiollm: Voice understanding and generation foundation models for natural interaction between humans and llms},
  author={An, Keyu and Chen, Qian and Deng, Chong and Du, Zhihao and Gao, Changfeng and Gao, Zhifu and Gu, Yue and He, Ting and Hu, Hangrui and Hu, Kai and others},
  journal={arXiv preprint arXiv:2407.04051},
  year={2024}
}

@inproceedings{graves2006connectionist,
  title={Connectionist temporal classification: labelling unsegmented sequence data with recurrent neural networks},
  author={Graves, Alex and Fern{\'a}ndez, Santiago and Gomez, Faustino and Schmidhuber, J{\"u}rgen},
  booktitle={Proceedings of the 23rd international conference on Machine learning},
  pages={369--376},
  year={2006}
}

@inproceedings{he2016deep,
  title={Deep residual learning for image recognition},
  author={He, Kaiming and Zhang, Xiangyu and Ren, Shaoqing and Sun, Jian},
  booktitle={Proceedings of the IEEE conference on computer vision and pattern recognition},
  pages={770--778},
  year={2016}
}

@article{irfan2021deepship,
  title={DeepShip: An underwater acoustic benchmark dataset and a separable convolution based autoencoder for classification},
  author={Irfan, Muhammad and Jiangbin, ZHENG and Ali, Shahid and Iqbal, Muhammad and Masood, Zafar and Hamid, Umar},
  journal={Expert Systems with Applications},
  volume={183},
  pages={115270},
  year={2021},
  publisher={Elsevier}
}

@article{santos2016shipsear,
  title={ShipsEar: An underwater vessel noise database},
  author={Santos-Dom{\'\i}nguez, David and Torres-Guijarro, Soledad and Cardenal-L{\'o}pez, Antonio and Pena-Gimenez, Antonio},
  journal={Applied Acoustics},
  volume={113},
  pages={64--69},
  year={2016},
  publisher={Elsevier}
}

@article{loshchilov2017decoupled,
  title={Decoupled weight decay regularization},
  author={Loshchilov, Ilya and Hutter, Frank},
  journal={arXiv preprint arXiv:1711.05101},
  year={2017}
}

@article{ji2025underwater,
  title={Underwater Target Recognition Method Based on Singular Spectrum Analysis and Channel Attention Convolutional Neural Network},
  author={Ji, Fang and Lu, Shaoqing and Ni, Junshuai and Li, Ziming and Feng, Weijia},
  journal={Sensors},
  volume={25},
  number={8},
  pages={2573},
  year={2025},
  publisher={MDPI}
}

@article{chen2024ship,
  title={A ship-radiated noise classification method based on domain knowledge embedding and attention mechanism},
  author={Chen, Lu and Luo, Xinwei and Zhou, Hanlu},
  journal={Engineering Applications of Artificial Intelligence},
  volume={127},
  pages={107320},
  year={2024},
  publisher={Elsevier}
}

@article{yao2024mobile_vit,
  title={Mobile\_ViT: underwater acoustic target Recognition Method based on local--global feature fusion},
  author={Yao, Haiyang and Gao, Tian and Wang, Yong and Wang, Haiyan and Chen, Xiao},
  journal={Journal of Marine Science and Engineering},
  volume={12},
  number={4},
  pages={589},
  year={2024},
  publisher={MDPI}
}

@article{zhao2025enhancing,
  title={Enhancing Underwater Acoustic Target Recognition Through Advanced Feature Fusion and Deep Learning},
  author={Zhao, Yanghong and Xie, Guohao and Chen, Haoyu and Chen, Mingsong and Huang, Li},
  journal={Journal of Marine Science and Engineering},
  volume={13},
  number={2},
  pages={278},
  year={2025},
  publisher={MDPI}
}

\end{document}